# Bulk transport properties of Bismuth selenide thin films approaching the two-dimensional limit

Yub Raj Sapkota, Dipanjan Mazumdar

Physics Department, Southern Illinois University, Carbondale, IL 62901

## ABSTRACT

We have investigated the transport properties of topological insulator $Bi_2Se_3$ thin films grown using magnetron sputtering with an emphasis on understanding the behavior as a function of thickness. We show that thickness has a strong influence on all aspects of transport as the two-dimensional limit is approached. Bulk resistivity and Hall mobility show disproportionately large changes below 6 quintuple layer which we directly correlate to an increase in the bulk band gap of few-layer $Bi_2Se_3$, an effect that is concomitant with surface gap opening. A tendency to crossover from a metallic to an insulating behavior in temperature-dependent resistivity measurements in ultra-thin $Bi_2Se_3$ is also consistent with an increase in the bulk band gap along with enhanced disorder at the film-substrate interface. Our work highlights that the properties of few-layer $Bi_2Se_3$ are tunable that may be attractive for a variety of device applications in areas such as optoelectronics, nanoelectronics and spintronics.

# INTRODUCTION

Topological insulators (TIs) are a new class of matter with bulk insulating states combined with metallic surface states that are topologically protected [1] [2]. Experimental realization of 3D topological states at room temperature and in absence of an external magnetic field has widened the interest of such materials beyond condensed matter physics [3-6]. Apart from applications in spintronics and fault-tolerant quantum computing [7-11], topological states are now considered for a variety of applications such as interconnects [12] and low-power electronics [13]. In this respect, the useful material properties must survive various tests of scalability and compatibility. Therefore, understanding the properties of TIs in the two-dimensional limit will provide the platform for future investigations regarding their applications.

Bismuth selenide ($Bi_2Se_3$) is the prototypical topological insulator material. Discovery of single Dirac cone [4] at the Brillouin zone center in $Bi_2Se_3$ has fueled tremendous interest in the research community. It is also a recognized thermoelectric material with a relatively small band gap (0.3 eV in bulk form). Therefore, $Bi_2Se_3$ is an ideal system for exploratory investigations into the application of TIs. Various thin-film growth techniques can be employed to grow TI materials. Molecular-beam epitaxy (MBE) has been widely used many groups [14-19]. Other methods include Pulsed laser deposition (PLD) [20], chemical vapor deposition [21] and magnetron sputtering [22]. Each method has its own advantages and disadvantages. From an application point of view, magnetron sputtering is versatile and less costly compared to PLD and MBE. Also, while not exclusive, sputtering is a well-established method for growing a variety of heterostructures in applied areas such as spintronics. Therefore, investigation of TI materials grown with magnetron sputtering is quite relevant.

Investigation of transport properties are of paramount importance to properly utilize the conducting surface states for device applications. However, both single crystal and thin-film $Bi_2Se_3$ suffers from high bulk carrier concentration, irrespective of growth method, that overwhelm the surface states (see Table S1 for a comparison on thin-film properties), and also can be induced by mechanical exfoliation [23], and exposure to ambient conditions [24] [25]. Studies show that chemical doping [26] [27], electrochemical doping [23] and synthesis in Se-rich conditions [28] can lower the bulk carrier concentration in $Bi_2Se_3$. But, as many recent investigations reveal, high carrier concentration samples do not preclude investigations and application of TIs. Recent reports demonstrated that heterostructures with high carrier concentration TIs show novel properties that is attributed to the unique nature of the surface states [29-32].

This work is focused on investigating the transport properties of $Bi_2Se_3$ thin films grown using radio frequency magnetron sputtering method with a particular emphasis on understanding the behavior of physical quantities such as concentration, resistivity and mobility as we approach the two-dimensional limit (few-layer properties) . The structure-property relationship of thin-films fabricated over a wide thickness range (2-100 quintuple layer) are discussed. Hall measurements reveal high bulk carrier concentration in all films, irrespective of thickness and substrate. We find that the bulk resistivity and Hall mobility values show disproportionately large variation with thickness as the 2D limit is approached. Leveraging on our prior work, where we demonstrated optical blue-shift in few-layer $Bi_2Se_3$ compared to bulk [33], we correlate the enhanced resistivity (and reduced mobility) behavior directly to the increase in the bulk band gap of $Bi_2Se_3$ below six quintuple layers (QL). Substrate also affects the transport properties in the few-layer regime through film-substrate disorder. Compared to isostructural $Al_2O_3$, the conductivity and mobility values of few-layer $Bi_2Se_3$ on amorphous quartz are lower.

We report a largely temperature-independent Hall coefficient and mobility response, and metallic resistivity behavior in 80-300K range. Overall, our work clearly explains several interesting properties of $Bi_2Se_3$ for device applications.

## EXPERIMENTAL DETAILS

$Bi_2Se_3$ was grown using commercially available stoichiometric target (Kurt Lesker, 99.999% purity) and RF sputtered in a high vacuum magnetron sputtering system (base pressure $4 \times 10^{-9}$ Torr). The growth rate for $Bi_2Se_3$ was over 10-15 QL/min even under moderate sputtering power conditions. Films were grown at room temperature and annealed *in-situ* at 300 C. $Bi_2Se_3$ thin films were fabricated in the 2-100 nm thickness range that roughly translates to 2-100 quintuple layers (1 QL ~ 0.95 nm). Structural analysis was performed on substrates such as $Si/SiO_2$ (100 nm), amorphous BN-buffered Si, 00*l*-oriented Si with a native $SiO_2$ layer, and c-oriented $Al_2O_3$ (sapphire), and transparent amorphous quartz ($SiO_2$). Structural and interface properties were characterized by means of high-resolution X-ray diffraction and reflectivity using a Rigaku Smartlab Diffractometer equipped with a Ge (220) 2-bounce incident beam monochromator to obtain a Cu K$\alpha_1$ radiation.

Transport properties were evaluated mainly on amorphous quartz ($SiO_2$) substrates to understand the performance of $Bi_2Se_3$ on $Si/SiO_2$, the semiconductor industry standard. Quartz substrates were chosen instead of $Si/SiO_2$ to prevent the influence of doped-Si substrate on transport. Also, in a previous study we evaluated the band gap properties of few-layer $Bi_2Se_3$ deposited on optically transparent quartz [33]. Therefore, the same conditions were used to evaluate transport. Transport (longitudinal resistivity, $\rho_{xx}$) and Hall measurements were performed using a NanoMagnetics Instruments ezHEMS system in the Van-der-Pauw (VDP) geometry. Measurements were performed on exposed $Bi_2Se_3$ films of various thickness deposited on 1x1cm substrates. To minimize surface contamination and aging effect, transport measurements were performed immediately after thin-film deposition. New samples were fabricated whenever necessary to improve the accuracy of our measurements. Four-probe ohmic contacts were made by making light contacts with Au/Cr probes onto silver paint deposited at the corners of the square sample. The measurements were completely automated using a LABVIEW program. Using VDP method, a vast range of parameters such as carrier-type, 2D and 3D carrier concentration, bulk resistivity, and Hall mobility were analyzed as a function of thickness and temperature. Additional complementary investigations were performed on $Bi_2Se_3$ films deposited on sapphire substrate to compare the transport properties on quartz.

## RESULTS AND DISCUSSION

In Figure 1a, we show the X-ray reflectivity (XRR) pattern of approximately 10-12 QL $Bi_2Se_3$ films grown on different substrates (Si, Si/SiO$_2$, amorphous BN, and c-orientated $Al_2O_3$). XRR is simultaneously a surface and bulk probe technique. An oscillatory pattern is observed in all cases that indicates a sharp interface with the substrate. Analysis of the critical angle reveals that the measured film densities are higher than the bulk value of 6.8 g/cm$^3$ by about 15-20 %.

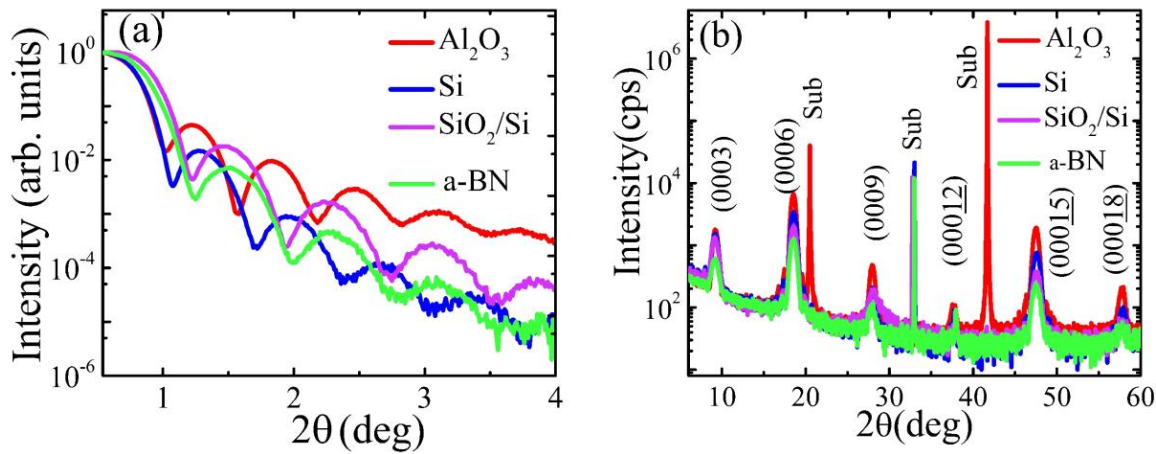

Figure 1(a)X-ray reflectivity of $Bi_2Se_3$ films deposited on different substrates as indicated. a-BN substrate implies amorphous BN grown on silicon substrate. (b) X-ray diffraction patterns of 10-12 nm Bi2Se3 films grown on different substrates. "Sub" indicated peaks from various substrates.

Though this value may be within margin of error in our measurement, it might also indicate the formation of a thin layer of $Bi_2O_3$ at the surface which has higher density than $Bi_2Se_3$.

Film roughness values are different from XRR data. Fits show that the film on $Al_2O_3$ is virtually atomically smooth while the film on $Si/SiO_2$ has 0.45 nm roughness. The film on bare Si (with a native oxide layer) and a-BN substrate show higher roughness (0.9 and 0.7 nm respectively) because of higher substrate roughness. The differences are within the error margin of the reflectivity fits.

X-ray diffraction analysis shows that films on $Al_2O_3$ possess the best crystallographic structure closely followed by $Si/SiO_2$. In Fig.1(b), we show the high-resolution $\theta$-$2\theta$ scans of the films shown in Fig 1(a). Only (000$l$) $Bi_2Se_3$ peaks are observed on all substrates apart from substrate peaks (indicated by "Sub") indicating out-of-plane (c-axis) growth in all cases. The film grown on $Al_2O_3$ (red) shows the highest intensity among all substrates whereas the film on amorphous BN (green) shows the lowest XRD intensity (by almost a factor of five for the (0006) peak). Film on $SiO_2/Si$ and Si are of intermediate quality. Similar characteristics are also inferred from the full-width-at-half-maxima (FWHM) values. There is no significant difference in the out-of-plane lattice parameter that indicates that strain effect is not important for $Bi_2Se_3$. Taken together with X-ray reflectivity data, it is clear that that $Bi_2Se_3$ films grown on $Al_2O_3$ have the best topographic and structural quality. The structural quality on amorphous substrates are also encouraging and we shall discuss the thickness evolution of $Bi_2Se_3$ structure on $Si/SiO_2$.

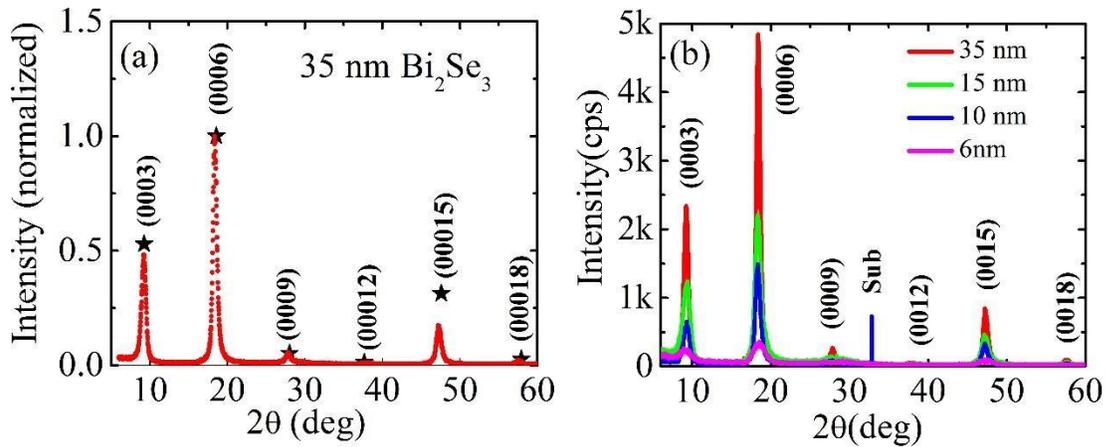

Figure 2 (a) Normalized XRD pattern of a 35 nm $Bi_2Se_3$ film grown on $Si/SiO_2$ showing only (000*l*) Bragg peaks. Simulated XRD intensities corresponding to (000*l*) Bragg peaks are shown using black stars. A very good agreement is observed except for the (00015) peak (b) XRD scans of $Bi_2Se_3$ thin films for different thicknesses as indicated. 'Sub" indicates substrate peak.

Figure 2a shows the high resolution θ-2θ XRD pattern of a 35 nm $Bi_2Se_3$ film deposited on $Si/SiO_2$ substrate. Various (000*l*) peaks assigned to $Bi_2Se_3$ are clearly observed. The intensities are normalized with respect to the highest (0006) peak in order to compare the data to a simulated XRD pattern [29] of bulk $Bi_2Se_3$ (c=28.63 Å) [35] . The (000*l*) simulated peak intensities are indicated by black stars. As evident, apart excellent agreement in the Bragg angle, the experimental intensities match simulation very well for all except the (00015) peak. This indicates that highly-oriented $Bi_2Se_3$ films are growing strain-free with a bulk-like crystal structure. The oriented crystal structure is also observed over a wide thickness range. In Fig. 2b, we show the XRD scans of films with different thickness deposited on $SiO_2/Si$. Only (000*l*) peaks are observed down to 6 QL. It is reasonable to assume that the structural integrity is maintained in films lower than 6 QL thickness. In supplementary figures S1-S3, we plot the XRD data on 4 nm $Bi_2Se_3$ deposited on sapphire and amorphous quartz substrate. Encouraged by the excellent structural properties observed in $Bi_2Se_3$ on amorphous substrates we proceed to discussing their transport properties.

Table 1. Carrier concentration (bulk and surface), bulk resistivity, and Hall mobility at 295 K of Bi$_2$Se$_3$ thin films deposited on quartz substrate, along with their optical bandgap as reported in ref [33]. N-type behavior was observed in all films.

| Thickness (nm) | Sheet concentration (cm$^{-2}$) | Bulk concentration (cm$^{-3}$) | Sheet resistance (h/e$^2$ ohm/sq) | Longitudinal Resistivity (ohm.cm) | Hall Mobility (cm$^2$/Vs) | Band Gap (eV) |
|---|---|---|---|---|---|---|
| 90 | 4.2×10$^{14}$ | 4.72×10$^{20}$ | 4.66×10$^{-3}$ | 1.08×10$^{-3}$ | 122.02 | - |
| 25 | 2.46×10$^{14}$ | 9.83×10$^{19}$ | 9.50×10$^{-3}$ | 6.1×10$^{-4}$ | 103 | 0.46* |
| 15 | 1.64×10$^{14}$ | 1.09×10$^{20}$ | 2.09×10$^{-2}$ | 8.10×10$^{-4}$ | 70.45 | 0.47 |
| 6 | 9.76×10$^{13}$ | 1.63×10$^{20}$ | 1.00×10$^{-1}$ | 1.56×10$^{-4}$ | 24.61 | 0.52 |
| 4 | 6.02×10$^{13}$ | 2.01×10$^{20}$ | 2.58×10$^{-1}$ | 2.00×10$^{-3}$ | 15.54 | 0.58 |
| 3 | 3.11×10$^{13}$ | 1.56×10$^{20}$ | 5.71×10$^{-1}$ | 2.95×10$^{-3}$ | 13.60 | 0.63 |
| 2 | 1.80×10$^{13}$ | 9.02×10$^{19}$ | 7.24 | 3.74×10$^{-2}$ | 1.86 | 0.82 |

*Band gap data on a 30 nm film

In table 1, we report the room temperature values of various transport properties for films grown on quartz substrates along with their optical band gap reported previously by the authors [33]. A close inspection reveals that thickness has a strong impact on all aspects of transport, particularly below 6 QL. Sheet carrier concentration (n$_{2D}$) shows two distinct regimes. Between 90-15 QL there is a reduction in n$_{2D}$ by about a factor of nearly 2.5. Below 6 QL, we notice a much sharper reduction in n$_{2D}$ values by roughly an order of magnitude as the thickness approaches the two-dimensional limit. This is consistent with the reports of Liu *et al*. [36] where the results are attributed to strong electron delocalization and topological protection. The bulk carrier concentration (n$_{3D}$) is largely insensitive to thickness and remains in the 10$^{19}$-10$^{20}$ range throughout. This again points to the fact that the bulk carrier concentration is dominated by chemical factors such as selenium vacancies.

The strongest effect of thickness is on the sheet resistance/longitudinal resistivity and Hall mobility values where overall over two order of magnitude change is observed. Normalized in terms of h/e$^2$=R$_k$=25.812kΩ, the Von-klitzing constant, the sheet resistance progressively increases from $4.6 \times 10^{-3}$ $R_k$ for the 90 QL film to 0.1 $R_k$ at 6 QL and crosses $1.0 R_k$ at the 2QL level. It is noteworthy to mention that the values reported here (sputtered films grown on quartz substrate) are comparable to MBE-grown $Bi_2Se_3$ on sapphire substrate as listed in Table S1 [28-32, 37, 38].

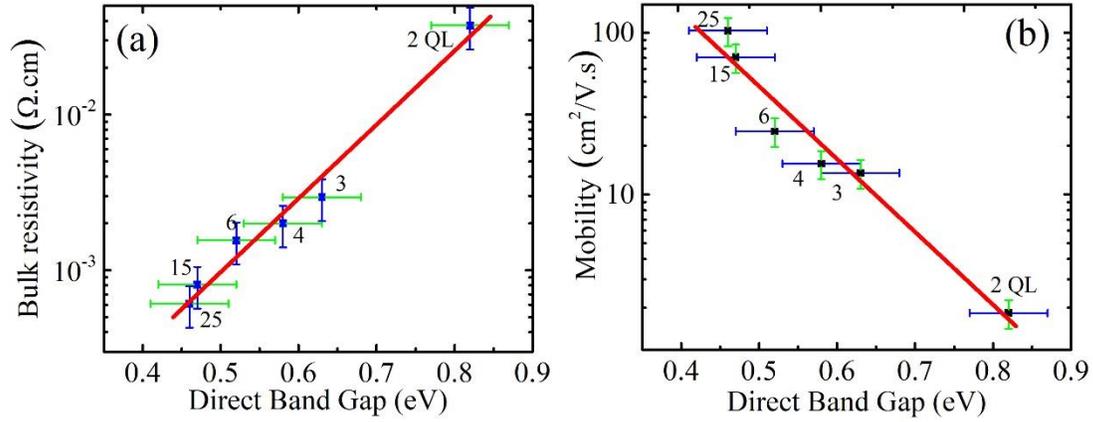

Figure 3 Variation of (a) bulk resistivity (b) Mobility in $Bi_2Se_3$ with measured optical band gap. An exponential behavior is observed. Optical band gap data reported in ref [33]. Each data point corresponds to a particular film thickness as indicated.

In Figure 3a, we plot the bulk longitudinal resistivity ($\rho_{xx}$) of films in the 2-25 QL thickness range along with the measured optical band gap of such films [33]. The optical absorption data and direct gap analysis is shown in Fig. S3. As clear from the figure, like the band gap behavior [33], the most drastic increase in transport is recorded below 6QL. We describe such disproportionately large variation directly to the blue-shift in the band gap of $Bi_2Se_3$ films as the two-dimensional limit is approached [33]. Previously, we noted that the $Bi_2Se_3$ bulk band gap blue-shifts by 0.3 eV as the thickness is reduced from 6QL to 2 QL. Such blue-shift has been verified in recent first-principle calculations [39] and is connected to the surface gap opening due to coupling of the top and bottom states. Taken together, we therefore find that bulk resistivity shows exponential growth with band gap, a trend indicative of semiconductor-like behavior. The observed scaling relation can be fitted to $\rho = e^{\beta E_g}$ ($E_g$=band

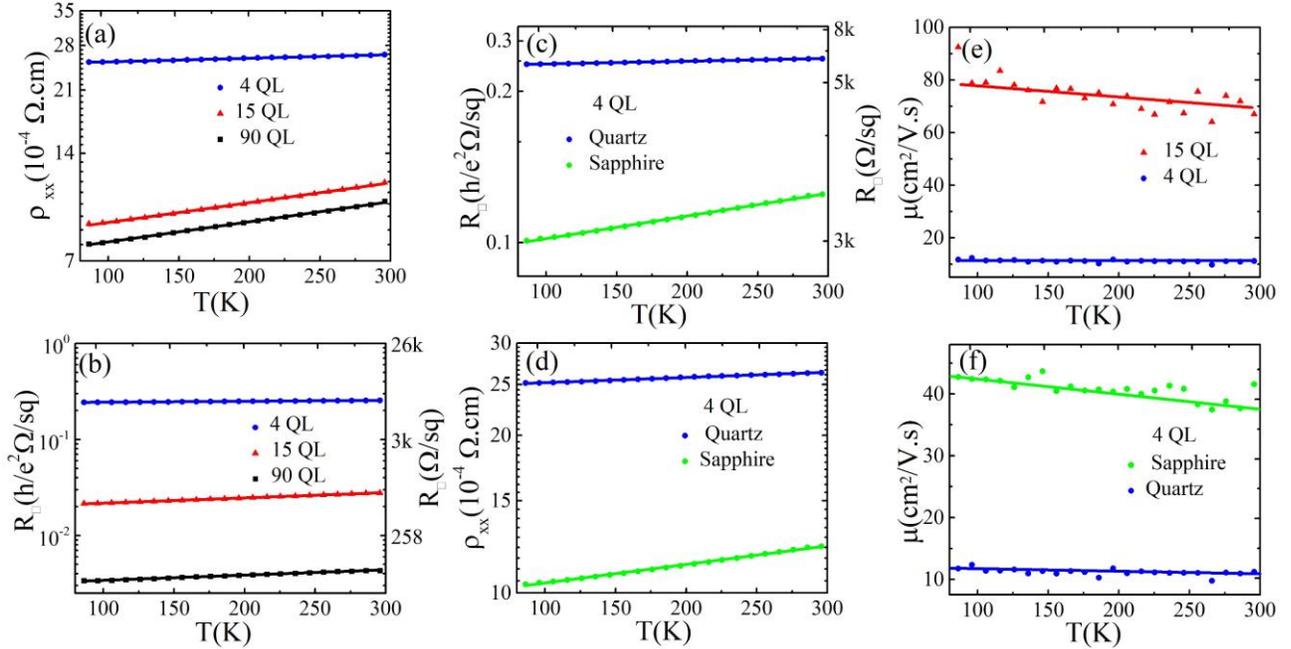

Figure 4 (a)Resistivity and (b) Sheet resistance of 4, 15, 90 QL Bi$_2$Se$_3$ deposited on quartz substrate. Comparison of (c) Resistivity and (d) Sheet resistance of a 4 QL film deposited on quartz and sapphire substrate. Sheet resistance is shown in terms of von-Klitzing constant and nominal value for comparison. (e) Hall mobility vs Temperature of 4 and 15 QL Bi$_2$Se$_3$ deposited on quartz substrate (f) Temperature variation of Hall mobility of a 4 QL film deposited on quartz and sapphire substrate.

gap) where β= 10.85±0.80, which is roughly equal to $(4k_BT)^{-1}$ and about half the semiconductor scaling factor of $(2k_BT)^{-1}$. The weaker than semiconductor behavior can be explained in terms of high carrier concentration in Bi$_2$Se$_3$ (degenerate semiconductor). We also note that the mobility reduces exponentially with band gap (Fig. 3b). Since Hall mobility is related to the resistivity through the relation $\mu = \frac{R_H}{\rho}$, ($R_H$ = Hall coefficient = 1/ne), the reduced mobility in few-layer Bi$_2$Se$_3$ can also be explained as due to the increase in band gap. This behavior is noted in semiconductors elsewhere, including graphene (a zero band gap system) where the mobility reduces dramatically when a band-gap is induced [40,41]. Therefore, our data in Fig. 3 proves conclusively that the behavior of ultra-thin Bi$_2$Se$_3$ films is sensitive to changes in the bulk electronic structure. Extrinsic factors such as carrier concentration and film-interface disorder also play a role as discussed next.

The bulk and sheet resistivity of 90, 15 and 4 QL Bi$_2$Se$_3$ film is shown as a function of temperature (85-295 K) in Fig 4 (a) and (b) respectively. Both sheet resistance and bulk resistivity show strong metallic tendency with a strong linear behavior ($d\rho_{xx}/dT$ is constant) in the entire temperature range. This has been previously attributed to strong electron-phonon scattering from high carrier concentration [42]. The slope of the $d\rho_{xx}/dT$ curve decreases as the thickness is reduced which is consistent with a crossover to insulating behavior reported in MBE-grown 2 and 3 QL films and attributed to scattering at the film-substrate [16,36]. We could not perform reliable transport measurements below 4 QL to verify this effect using the VDP method. While our data cannot rule out the influence of disorder at the film-substrate interface, but in light with the data reported in Fig. 3, we assert that the crossover tendency below 6 QL can be attributed to the intrinsic tendency of $Bi_2Se_3$ to blue-shift. It is worth emphasizing that both tendencies promote an insulating behavior, and it is not possible to distinguish their signatures from the temperature dependent behavior alone. However, if disorder effect was indeed stronger, then the optical blue-shift data will also be obscured, which is not the case (see Fig. S3 and ref [33]). Therefore, we assert that the change in band gap is the dominant effect causing the large changes in resistivity and mobility.

To understand the effect of interface disorder through the substrate, we have additionally fabricated and investigated $Bi_2Se_3$ films grown on sapphire substrate. The room temperature values of the physical quantities on sapphire substrate is reported in Table S2. In Fig 4 (c) and (d), we compare the temperature-dependence of bulk and sheet resistivity of a 4 QL (below 6 QL) $Bi_2Se_3$ film grown on quartz and sapphire substrate. Assuming that the bulk gap behavior is identical in the two samples, the difference in transport properties can be rationalized in terms of film-substrate interface. It is clearly observed in the data. The resistivity of the 4 QL film is higher (nearly double) at room temperature on quartz compared to sapphire. Also the metallic behavior, as inferred from the slope in $d\rho_{xx}/dT$ curve, is stronger on sapphire proving that inclination for insulating behavior is weaker. But as our data in Fig. 3 show, the biggest change in band gap is also below 3 QL, which coincides with enhanced interface disorder. However, the universal nature of the scaling behavior, irrespective of the growth method (MBE, sputtering) and substrate (amorphous, crystalline) points to an intrinsic behavior as reported here. We assert that the exponential dependence of resistivity/mobility will prevail (with possibly a smaller scaling constant), even when interface disorder is reduced

Both the changing electronic structure through thickness and interface-disorder from substrate have a strong effect on the temperature dependence of Hall mobility. In Fig. 4 (e), we show temperature dependence of Hall mobility measured using the van-der-Pauw method on 4 and 15 QL films on quartz substrate. The Hall mobility is virtually independent of temperature for the 4 QL film and changes only by 0.5 cm$^2$/V.s in the entire temperature range. A slightly stronger temperature dependence is observed in the 15 QL film. The insensitivity to temperature is attributed to the high carrier concentration in all the films [43]. In Fig. 4f, we compare the variation of Hall mobility in 4 QL Bi$_2$Se$_3$ deposited on sapphire to quartz substrate. The mobility is about 2-3 times higher on sapphire. This can be explained as due to increased conductivity of Bi$_2$Se$_3$ films on sapphire (Fig. 4 c and d) and lower carrier concentration on sapphire (see Table S2, Hall coefficient = 1/ne). However, even on sapphire substrate, Hall mobility is weakly dependent on temperature in the 85-300K range that is consistent with high carrier concentration values. Together with the resistivity measurements, we show that few-layer Bi$_2$Se$_3$ films are also sensitive to substrate-induced disorder and superior transport properties are observed on sapphire which provides a better lattice match and low interface disorder with Bi$_2$Se$_3$. The extrinsic factors affecting Bi$_2$Se$_3$ properties (disorder, carrier concentration) can be improved with a lattice-matched substrate and tuning the growth conditions. Once such advances are properly implemented, the transport properties of Bi$_2$Se$_3$ will be dominated by their intrinsic bulk and surface properties.

# CONCLUSION

Transport properties of $Bi_2Se_3$ thin-films fabricated using magnetron sputtering is investigated as a function of thickness. The trends in transport properties can be explained by a combination of intrinsic (electronic structure changes) and extrinsic (high carrier concentration, film-substrate disorder) factors. We show that the disproportionately large enhancement in bulk resistivity values, and strong reduction in Hall mobility, in films below 6 QL can be attributed to an increase in the bulk band gap of $Bi_2Se_3$ that occurs simultaneously with the emergence of gapped surface states the two-dimensional limit is approached. High carrier concentration in $Bi_2Se_3$ which leads to metallic resistivity behavior and temperature independent Hall behavior. Substrate induced film-substrate disorder also promotes an insulating behavior as the two-dimensional limit is approached. Overall, our work amply demonstrates that transport properties of $Bi_2Se_3$ films are highly tunable through finite-size effects. This can be of importance in its applications in areas such as optoelectronics, nanoelectronics and spintronics.

# ACKNOWLEDGEMENTS

DM would like to thank start-up funds from Southern Illinois university for support of this work. We would like to thank Dr. Sujit Singh of NanoMagnetics Instruments for help and advice on the transport measurements.